\documentclass[sigconf]{acmart}

\usepackage{booktabs} 
\usepackage{enumitem}
\usepackage{subcaption}
\usepackage[utf8]{inputenc}

\setcopyright{none}
\usepackage{xcolor}

\acmDOI{}

\acmISBN{}

\acmConference{arXiv}{2018}{} 
\acmYear{2018}
\copyrightyear{2018}
\acmPrice{}

\begin{document}
\title{Neuroscientific User Models}
\subtitle{The Source of Uncertain User Feedback and Potentials for Improving Web Personalisation}

\author{Kevin Jasberg}
\affiliation{%
  \institution{Web Science Group -- Heinrich-Heine-University}
  \city{Duesseldorf} 
  \state{Germany} 
  \postcode{45225}
}
\email{kevin.jasberg@uni-duesseldorf.de}

\author{Sergej Sizov}
\affiliation{%
  \institution{Web Science Group -- Heinrich-Heine-University}
  \city{Duesseldorf} 
  \state{Germany} 
  \postcode{45225}
}
\email{sizov@hhu.de}

\renewcommand{\shortauthors}{}

\begin{abstract}
In this paper we consider the neuroscientific theory of the Bayesian brain in the light of adaptive web systems and content personalisation. In particular, we elaborate on neural mechanisms of human decision-making and the origin of lacking reliability of user feedback, often denoted as noise or human uncertainty. To this end, we first introduce an adaptive model of cognitive agency in which populations of neurons provide an estimation for states of the world. Subsequently, we present various so-called decoder functions with which neuronal activity can be translated into quantitative decisions. The interplay of the underlying cognition model and the chosen decoder function leads to different model-based properties of decision processes. The goal of this paper is to promote novel user models and exploit them to naturally associate users to different clusters on the basis of their individual neural characteristics and thinking patterns. These user models might be able to turn the variability of user behaviour into additional information for improving web personalisation and its experience.
\end{abstract}

\maketitle

\section{Introduction}
\begin{figure}[t]
    \centering
    \begin{subfigure}{0.495\linewidth}
        \includegraphics[width=\textwidth]{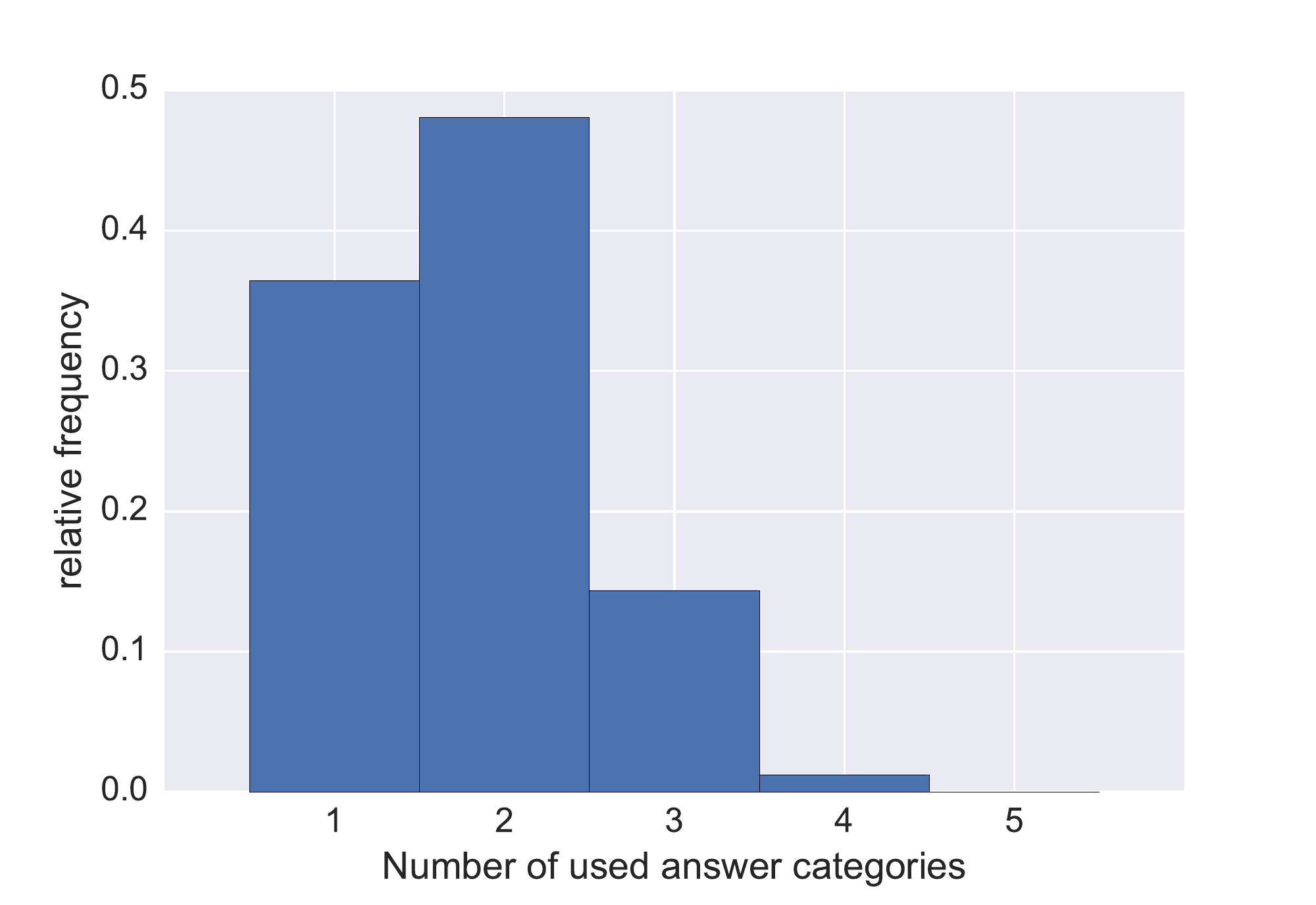}
        \caption{Frequency of use of response categories (opinion changes).}
         \label{fig:MotExa1}
    \end{subfigure}
     \hfill
    \begin{subfigure}{0.495\linewidth}
        \includegraphics[width=\textwidth]{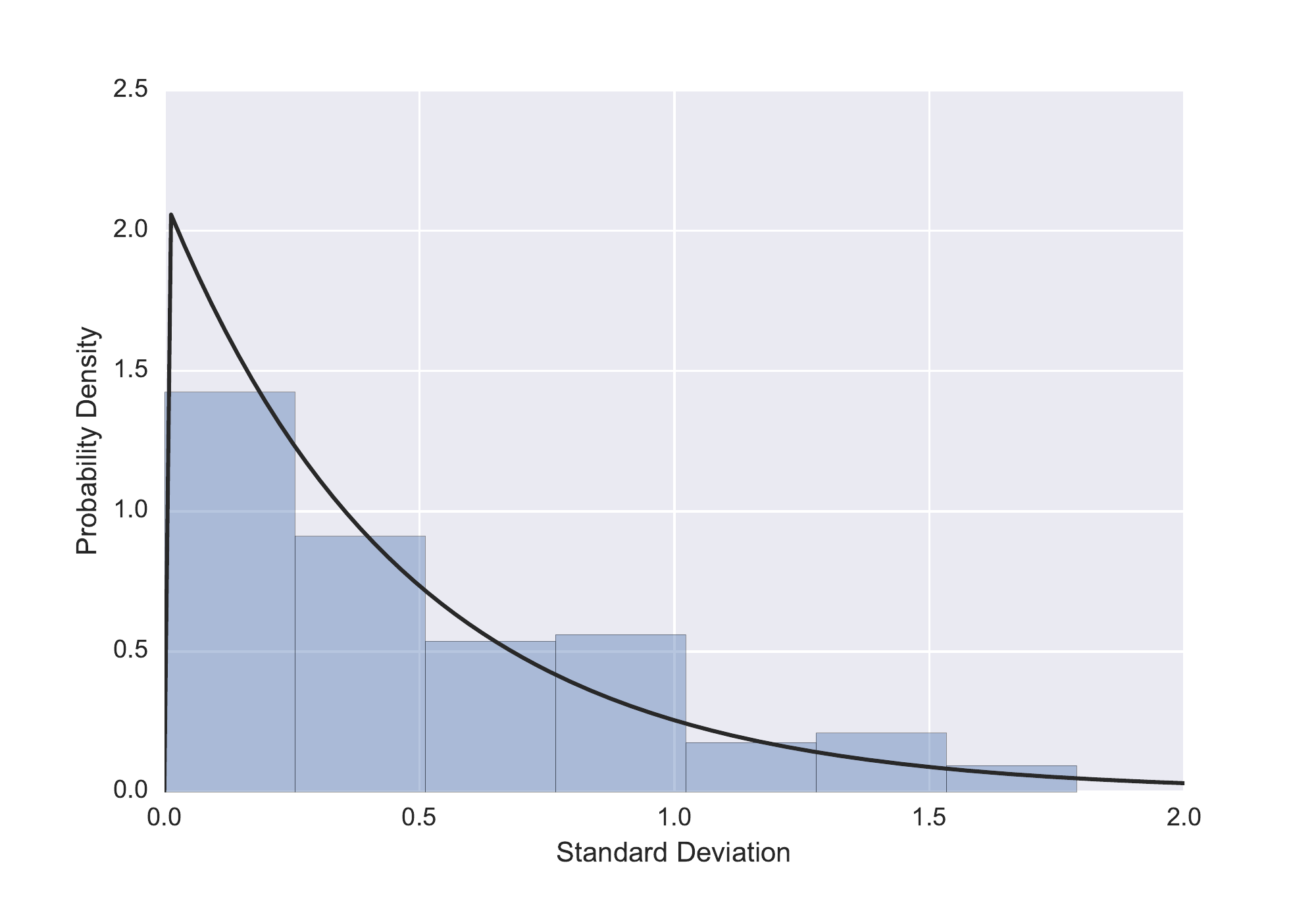}
        \caption{Distribution of user variances together with Pareto ML-fit.}
         \label{fig:MotExa2}
    \end{subfigure}
    \vspace*{-3mm}
    \caption{Evidence data for uncertain user responses gathered from five repeated trials of movie trailer ratings.}
    \label{fig:MotivatingExample}
    \vspace*{-5mm}
\end{figure}

Recent research has revealed that there is a considerable lack of reliability for user feedback when interacting with adaptive systems, often denoted as user noise or as human uncertainty \cite{RateAgain, JasUMAP}. This effect can be made visible through repeated feedback tasks, e.g. the rating of items. Figure \ref{fig:MotivatingExample} depicts the outcome of a self-conducted experiment in which participants had to (re-)rate same movie trailers five times with a small temporal gap between each rating trial. Subfigure \ref{fig:MotExa1} shows that only 35\% of all participants gave constant ratings, always using the same answer category on the same trailer. The remaining 65\% changed their rating on the same trailer several times. When considering user feedback as a random variable following a certain distribution, the variances can be seen as representations of feedback uncertainty. Subfigure \ref{fig:MotExa2} depicts the distribution of all those uncertainties in our experiment.

The idea of underlying distributions for user feedback is not far-fetched since renowned neuroscientists assume that any decision-making is based on an intern distribution encoded by neural activity \cite{Pouget} and updated through Bayesian inference in recurrent neural networks \cite{Friston}.
At any time when a decision has to be made, one has to consider a variety of yet unknown states of the world which are most relevant for the decision process itself. According to \cite{Friston}, each of these states are unconsciously estimated by an agent (population of neurons) and thus being made accessible to the brain. In doing so, there is evidence that each agent provides a probability density over possible values of such a state of the world (probabilistic populaton codes) and thus also accounts for the uncertainty of a quantity \cite{Pouget}. However, these estimations slightly differ in each cognition trial due to the volatile concentration of released neurotransmitters, impacting the spiking habits of downstream neurons (neural noise).

\begin{figure}[b]
    \centering
    \begin{subfigure}{0.325\linewidth}
        \includegraphics[width=\textwidth]{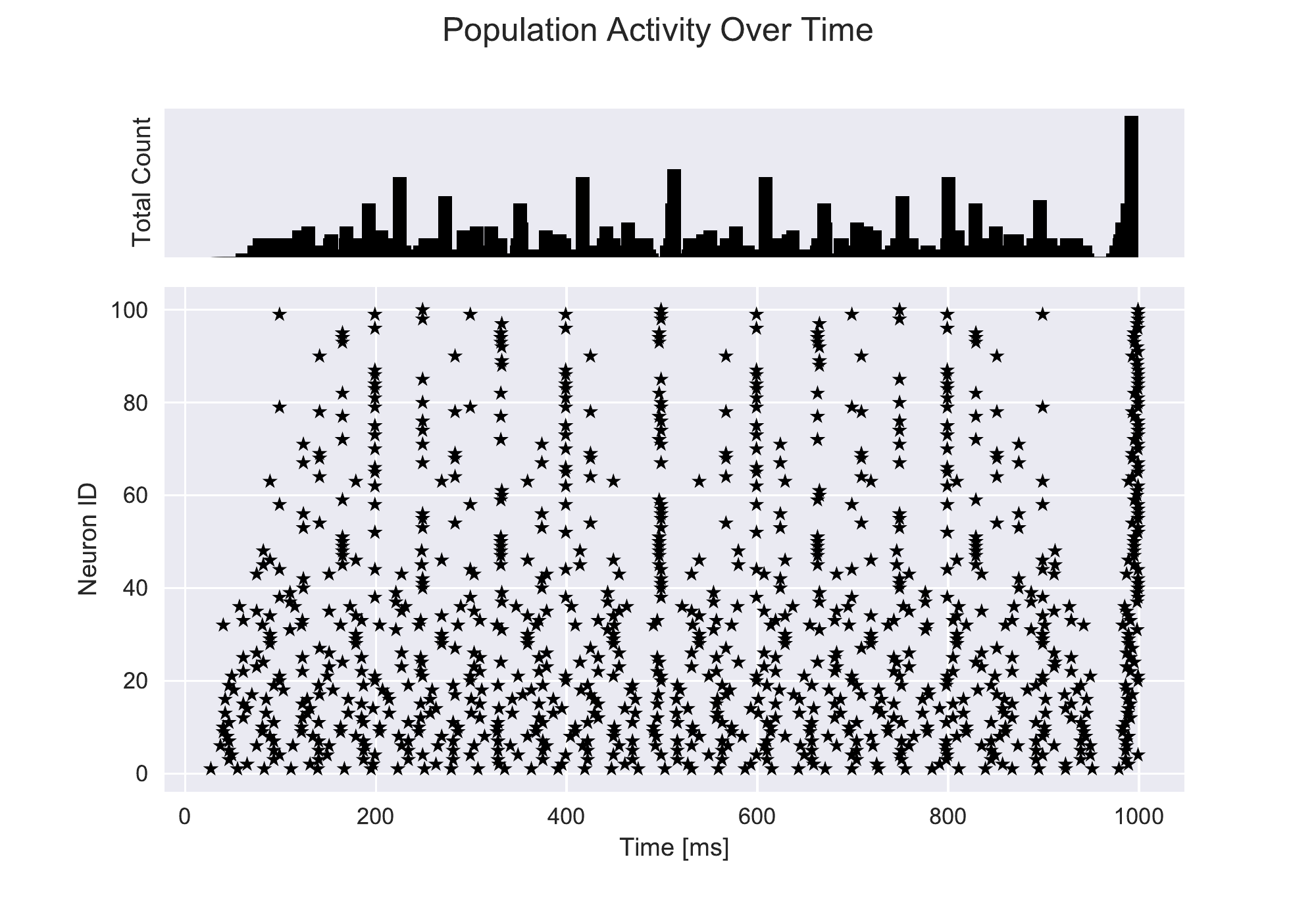}
        \caption{1 stars}
         \label{fig:PAOT1}
    \end{subfigure}
     \hfill
    \begin{subfigure}{0.325\linewidth}
        \includegraphics[width=\textwidth]{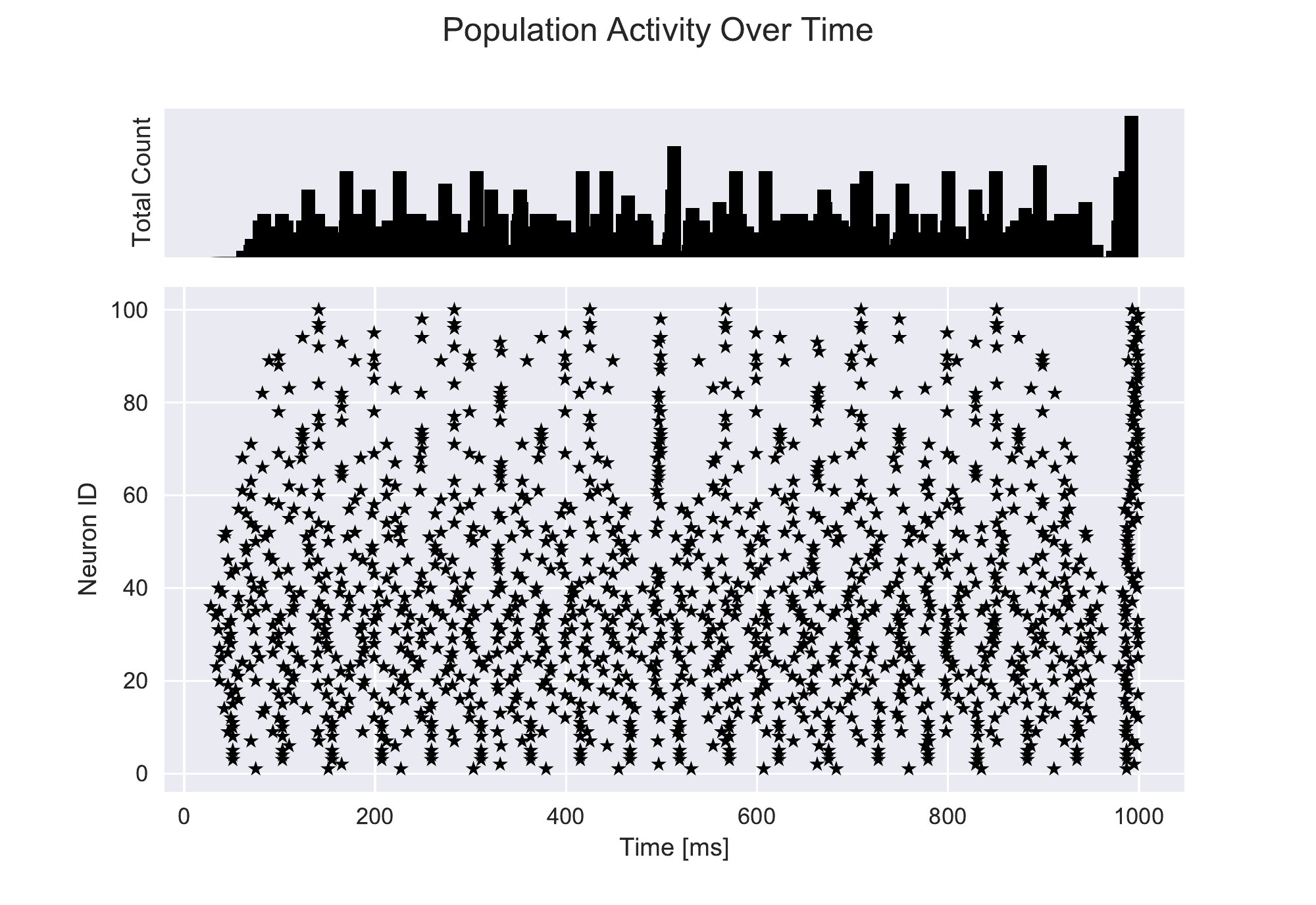}
        \caption{2 stars}
         \label{fig:PAOT2}
    \end{subfigure}
     \hfill     
    \begin{subfigure}{0.325\linewidth}
        \includegraphics[width=\textwidth]{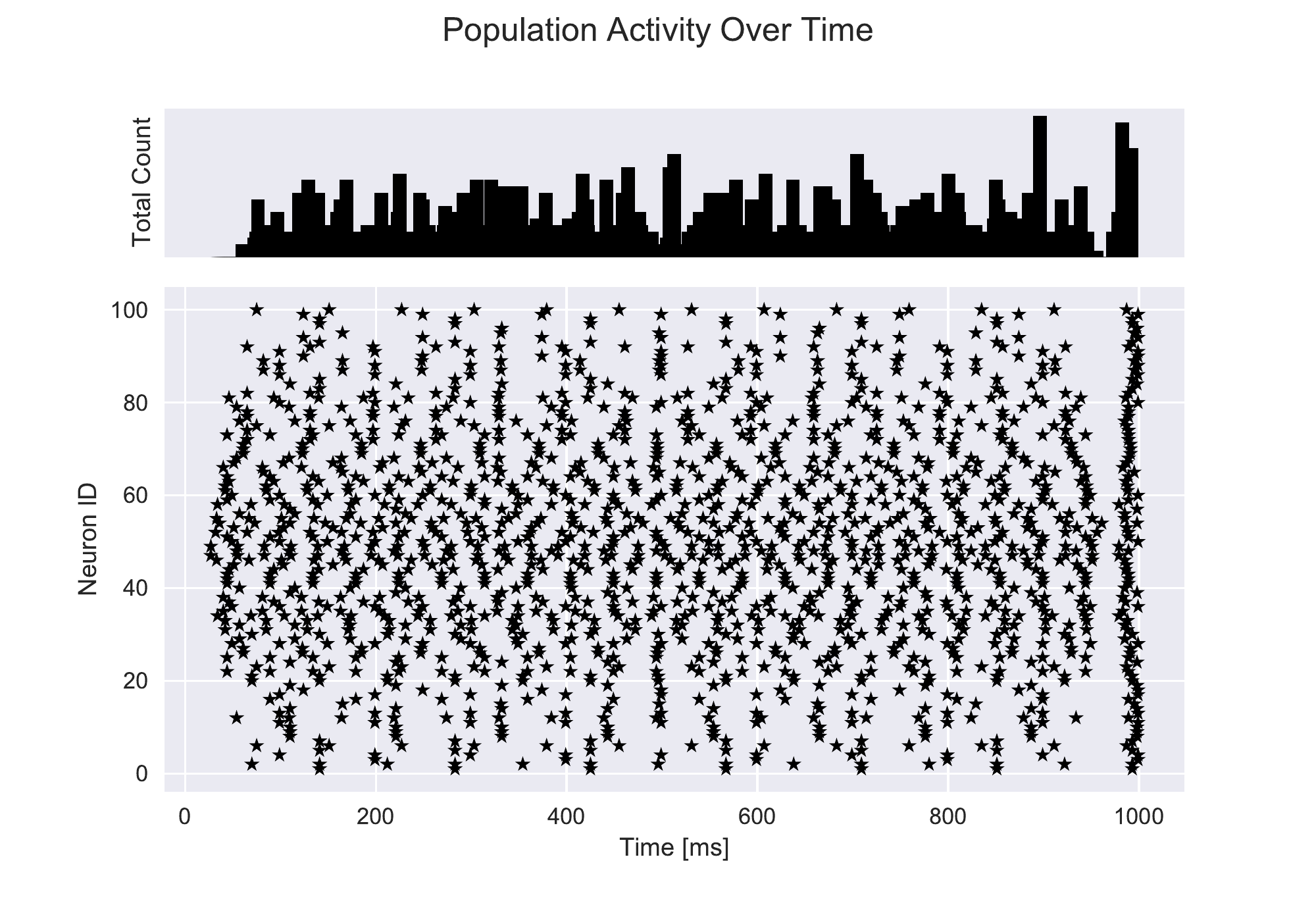}
        \caption{3 stars}
         \label{fig:PAOT3}
    \end{subfigure}
    \vspace*{-3mm}
    \caption{Raster plot of population activity encoding a specific rating on a 5-star scale by location and frequency.}
    \label{fig:PAOT}
    \vspace*{-1mm}
\end{figure}

\section{Framework}
\begin{figure*}
    \centering
    \begin{subfigure}{0.325\linewidth}
        \includegraphics[width=\textwidth]{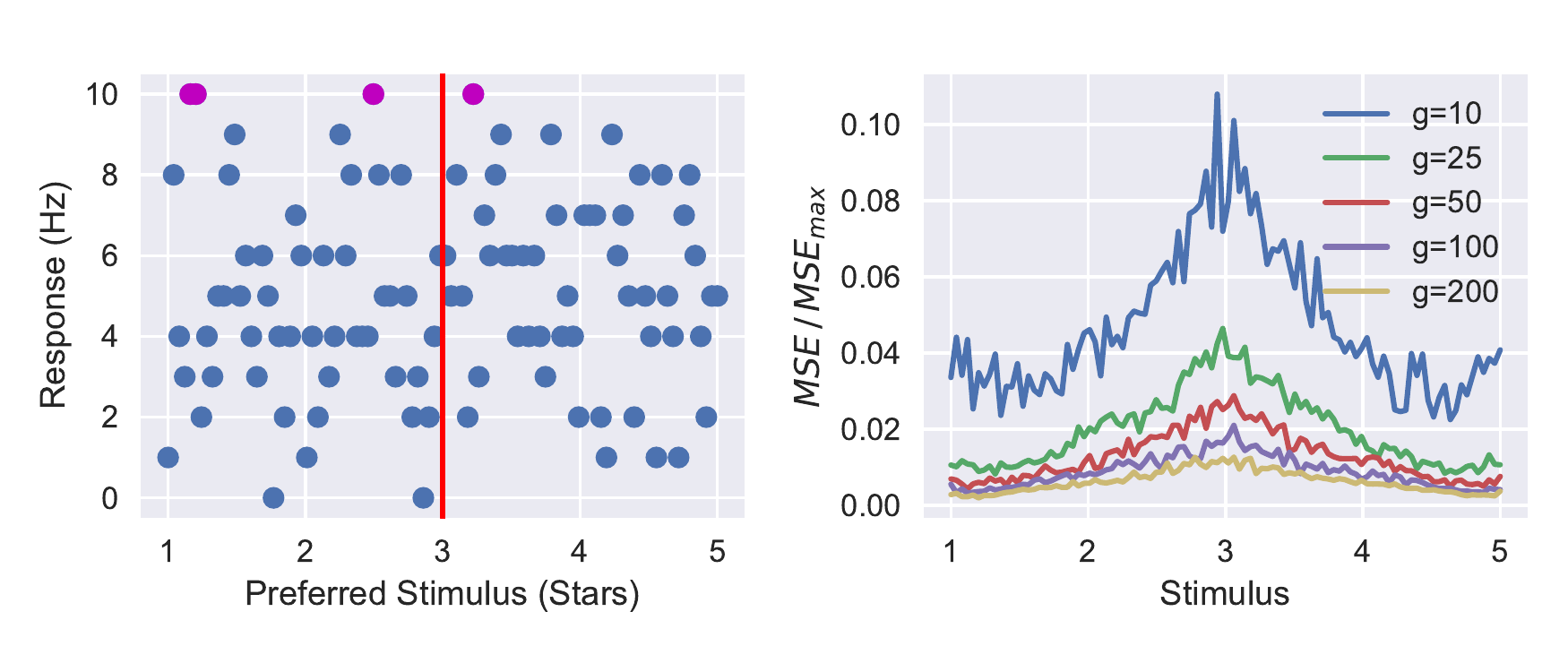}
        \caption{Mode Value Decoder}
         \label{fig:MVD}
    \end{subfigure}
     \hfill
    \begin{subfigure}{0.325\linewidth}
        \includegraphics[width=\textwidth]{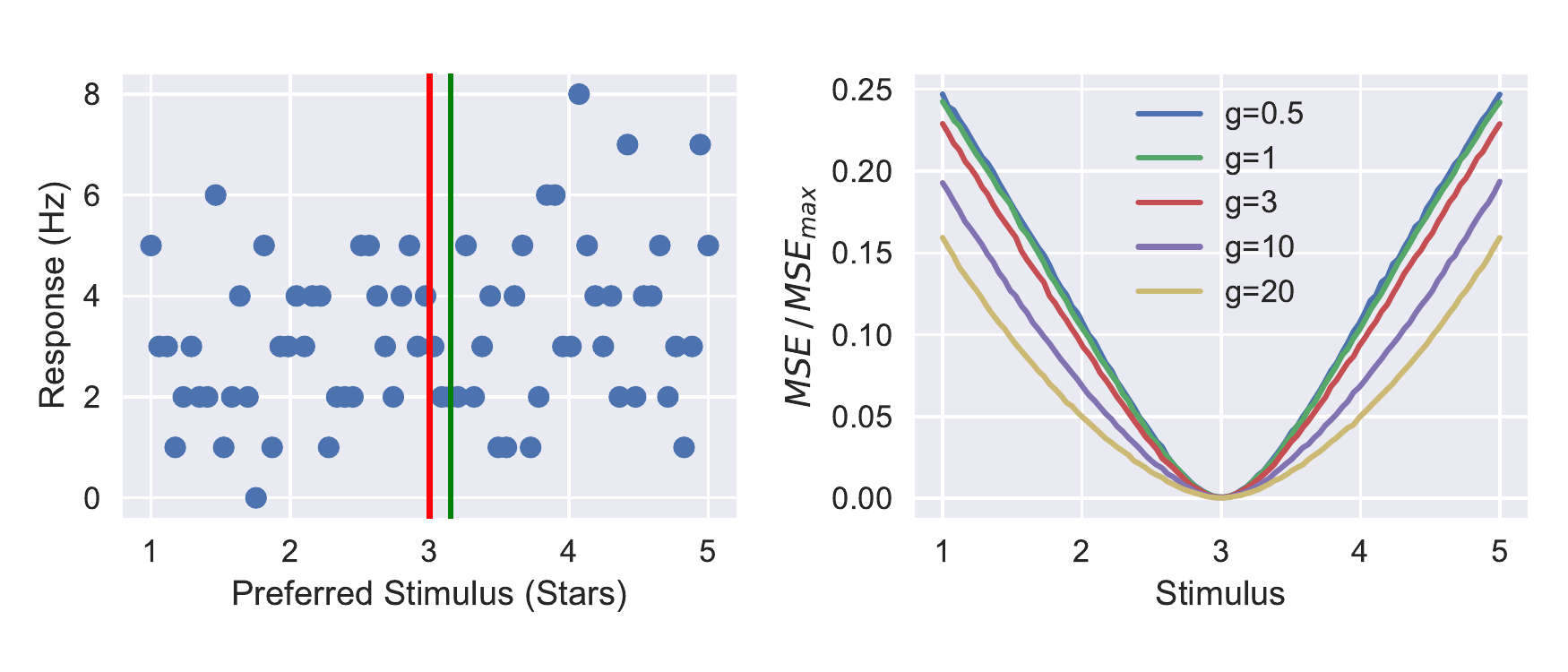}
        \caption{Weighted Average Decoder}
         \label{fig:WAD}
    \end{subfigure}
     \hfill     
    \begin{subfigure}{0.325\linewidth}
        \includegraphics[width=\textwidth]{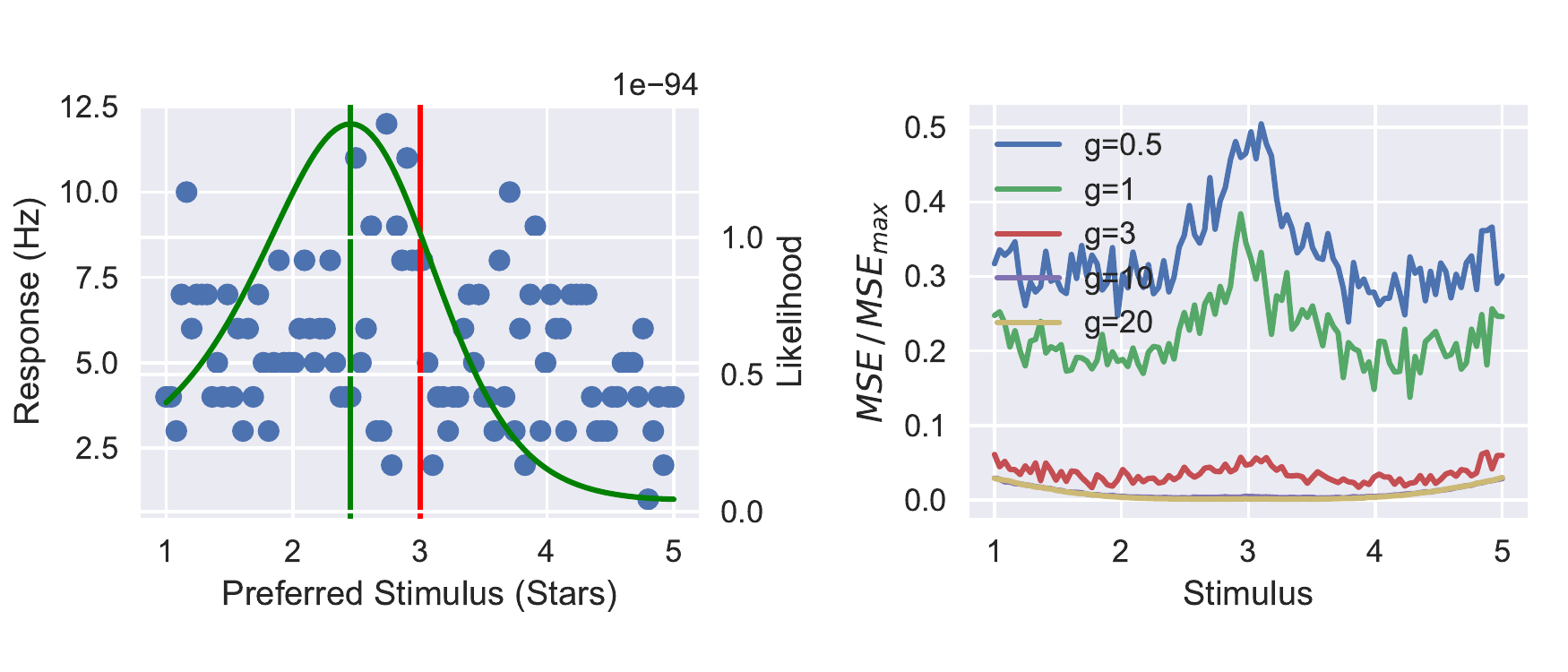}
        \caption{Maximum Likelihood Decoder}
         \label{fig:MLD}
    \end{subfigure}
    \vspace*{-3mm}
    \caption{Population response for a 3-star decision together with estimates obtained from different decoder functions}
    \label{fig:Decoder}
    \vspace*{-4mm}
\end{figure*}

\subsection*{The Single Neuron Model}
The response $r$ of a single neuron to a stimulus $S$ is limited to transmission of electric impulses (spiking) and since each neuron has only got two states of activation, theories of neural coding assume that information is encoded by the spiking frequency (rate) \cite{BayesianBrain}. The functional relationship between responses $r$ and the characteristics $s$ of a stimulus $S$ is given by the tuning curve $r=f(s)$. Higher cognitions are typically modelled by bell-shaped tuning curves, i.e. $f(s) = g\cdot h(s_p,w^2)(s) + f_0$, where the shape emerges from the Gaussian density function $h$ with mean $s_p$ and variance $w^2$ \citep{BayesianBrain}. The additional components $g,f_0\!\in\!\mathbb{R}$ represent a frequency gain and offset respectively. Each tuning curve maximises for a particular value $s_p:=\operatorname{argmax} f$, denoted as the preferred stimulus. According to \cite{Tolhurst}, these responses $r$ are subject to a variability and must therefore be seen as random variables $R$ rather than fixed values determined by tuning curves. It has been found that $R\!\sim\!\operatorname{Poi}(\lambda)$ follows a Poisson distribution with expectation $\lambda=f(s)$ \cite{Pouget}.
\vspace{-2mm}

\subsection*{Probabilistic Population Codes}
Let us consider a population with $N$ neurons. For a given stimulus (or state of the world) $s$, each neuron will respond with $r_i$ according to the distribution of $R_i$. Since the magnitude of these outcomes is strongly dependent on the underlying tuning curve $f(s)$ the population response $r=(r_1,\ldots,r_N)$ is a representation of $s$. 
This is exemplified in Fig. \ref{fig:PAOT} which depicts the population responses of different estimations for a rating on a 5-star scale. One can see that the decision is encoded by location and frequencies. Due to neural noise, these patterns slightly change from trial to trial.\vspace{-2mm}

\subsection*{Decoder Functions}
\paragraph*{Mode Value Decoder}
Due to the construction of tuning curves, the MVD assumes that it is exactly the neuron with maximum spiking frequency that is most likely to be addressed by the stimulus or the state of the world. The decoded estimation is therefore given as
$\hat{s} = s_{p,i}$ with $i = \operatorname{argmax} r_i$. Figure \ref{fig:MVD} depicts a population response for a 3-star-decision (red line) together with possible estimators (magenta dots) for this decision. There is a large ambiguity of estimators along with a heavy impact of repetitions which diminishes for higher frequencies in neural responses. Reliability analysis (via fraction of maxMSE) reveals that this decoder is suitable for users who show a low uncertainty for extreme ratings (1 or 5) and higher variability for middle ratings. This might apply to users who frequently choose these extremes and use middle ratings sparsely. 

\paragraph*{Weighted Average Decoder} The WAD accounts for all responses by setting the specific frequency $r_i$ as a weight to the corresponding preferred value $s_{p,i}$ and considers its contribution to the total response, i.e. 
$\hat{s} = (\sum r_i s_{p,i}) / (\sum r_i)$. Figure \ref{fig:WAD} shows a population response for a 3-star-decision (red line) together with the WAD estimator (green line). This estimator is not ambiguous and not heavily impacted by repeated decision-making. This decoder corresponds to users whose uncertainty minimises for middle ratings, i.e. users who often give moderate ratings and use extremes sparsely.

\paragraph*{Maximum Likelihood Decoder}
For a given population response, the MLD chooses $s$ with an eye to maximise the corresponding likelihood 
$P(r|s) = \prod f_i(s)^r_i/r_i! \,e^{-f_i(s)}$, i.e. $\hat{s} = \operatorname{argmax} P(r|s)$. In Fig. \ref{fig:MLD} we see another response for the same decision along with the Likelihood and the MLE (green line). The MLD is the first decoder that explicitly accounts for neural noise which makes it more variable for repetition than the previous ones. By frequency modulation, the MLD can account for users with high uncertainty for middle ratings and low uncertainty for extremes and  vice versa. Altogether, this decoder is appropriate for users who show high uncertainty in their decisions and utilise the whole scale.

\paragraph*{Maximum A Posteriori Decoder}
The likelihood can be transformed into a probability function over the stimulus via Bayes' theorem, i.e. $P(s|r) \propto P(r|s)P(s)$. $P(s)$ denotes prior belief about the stimulus or the states of world that has been learned through former experiences. The estimator is then chosen so that this posterior is maximised, i.e. $\hat{s} = \operatorname{argmax} P(s|r)$. The MAD is much like the MLD but with less variability since the prior works as stabiliser. Therefore, this decoder works well for users utilising the whole scale while having only moderate uncertainty in their decisions.

\vspace{-2mm}
\section{Discussion}
In this paper, we introduced neuroscientific user models representing thinking patterns by medical correlates. By adding neural noise as measured in vivo, these allow us to explicitly account for the variability of user feedback. 
The interaction of medical correlates with decoder functions makes these user models very adaptable.
Initial investigations show that each user of our experiment (outlined in the Introduction) can be described by an indvidual set of neural parameters together with a decoder function, i.e. uncertain user ratings can be well reproduced.
Therefore, these models hold great potential for web research, especially for personalisation and recommendation since they provide additional information about user behaviour which are missing from other models.
They also map uncertainty into a parameter space in which users can be clustered by their neural characteristics. In future research, we will address the quality of this clustering and continue to focus on the reproduction of uncertainty. This will lead to a new way of web  development/assessment with more respect for the human nature.
\vspace{-2mm}

\bibliographystyle{ACM-Reference-Format}
\bibliography{Literatur} \pagebreak


\end{document}